\documentclass[manuscript]{aastex}

\usepackage{amsmath}
\usepackage{amssymb}
\usepackage{graphics}

\def\TAME{{TAME}}
\def\teff{$T_\mathrm{eff}$}
\def\logg{log $g$}
\def\kms{$\mathrm{km~s^{-1}}$}
\def\ew{{\it EW}}
\def\ews{{\it EW}s}
\newcommand{\FeI}{\ion{Fe}{1}}
\newcommand{\FeII}{\ion{Fe}{2}}

\shorttitle{Tool for Automatic Measurement of Equivalent width (TAME)}
\shortauthors{Kang and Lee}

\begin{document}

\title{Tool for Automatic Measurement of Equivalent width (TAME) }

\author{Wonseok Kang$^{1,~2}$, Sang-Gak Lee$^2$} 
\affil{
       $^1$School of Space Research, Kyung Hee University, Yongin-Si, Gyeonggi-Do 446-701, Republic of Korea \\
       $^2$Astronomy Program, Department of Physics and Astronomy, Seoul National University, Seoul 151-742, Republic of Korea
      }

\begin{abstract}
We present a tool for measuring the equivalent width (\ew ) in high-resolution spectra. 
The Tool for Automatic Measurement of Equivalent width (\TAME) provides the \ews\ of spectral lines by profile fitting in the automatic or the interactive mode, which can yield a more precise result through the adjustment of the local continuum and fitting parameters.
The automatic \ew\ results of \TAME\ have been verified by comparing them with the manual \ew\ measurements by  IRAF {\tt splot} task using the high-resolution spectrum of the Sun, and measuring \ews\ in the synthetic spectra with different spectral resolutions and S/N ratios.
The \ews\ measured by \TAME\ agree well with manually measured values, with a dispersion of less than 2 m\AA.   
By comparing the input \ews\ for synthetic spectra and \ews\ measured by \TAME, we conclude that it is reliable for measuring the \ews\ in a spectrum with a spectral resolution, R $\gtrsim$ 20000 and find that the errors in \ews\ is less than 1 m\AA\ for a S/N ratio $\gtrsim$ 100. 
\end{abstract}


\section{Introduction}

The measurement of equivalent width (\ew) for spectral absorption lines is essential in a spectral analysis, particularly for determining the atmospheric parameters and chemical abundances of stars. 
In the study of stellar spectroscopy, it is critical to determine the atmospheric parameters of stars, such as the effective temperature (\teff), surface gravity (\logg), metallicity ([Fe/H]), and micro-turbulence ($\xi_t$), because atmospheric parameters are fundamental to understand spectroscopic properties and construct the model atmosphere for an abundance analysis.
For the atmospheric parameters, however, the most common method is to analyse the abundances that can be obtained from \ew\ measurements of neutral and singly ionized lines.
Additionally, the chemical abundances are also estimated by measuring the \ews\ of atomic lines  \citep[e.g.,][]{bensby03, santos04, bond06, gilli06, sousa06, kang11}.
The \ew\ measurement, therefore, is undoubtedly the most important task in spectroscopic studies.

The \ews\ of spectral lines have generally been measured by using the {\tt splot} task in IRAF\footnote{IRAF is the Image Reduction and Analysis Facility software. It is written and supported by the IRAF programming group at the National Optical Astronomy Observatories (NOAO) that is operated by the Association of Universities for Research in Astronomy (AURA), Inc., under cooperative agreement with the National Science Foundation} {\tt echelle} package, which makes it possible to manually estimate the \ew\ of each line.
Although this method guarantees a high degree of accuracy for \ew\ measurement, it requires a disciplined expert in the field of stellar spectroscopy and the result depends on the personal bias.
For an abundance analysis, it is necessary to measure the \ews\  for many lines for each star, which is a tedious and time-consuming task. 
Therefore, a uniform and fast method for \ew\ measurement is required in stellar abundance studies using a large number of high-resolution spectra. 
 
\citet{ARES} presented a new C++ code, called ARES (Automatic Routine for line Equivalent widths in stellar Spectra), which can automatically and simultaneously measure the \ews\ of spectral lines in stellar spectra. 
ARES provides quick measurement results for \ews, without manual operation, from high-resolution spectra.
However, ARES code focuses on the performance of the code, and hence deprives a user of the interactive operation that can be used to control an environment for each line. 
Further, a Fortran code for \ew\ measurement, called DAOSPEC, was recently presented by \citet{DAOSPEC}. 
In order to achieve more accurate measurement, DAOSPEC offers the enhanced interactive mode for detailed manipulative tasks, such as the adjustment of the local continuum and the deblending of nearby lines. 
Unfortunately, ARES and DAOSPEC were written in C++ and Fortran, respectively.
Therefore, the installation of ARES and DAOSPEC depends on the platform OS, and it would be difficult and inconvenient to compile and run these codes coherently, because of the required libraries (e.g., cfitsio\footnote{http://heasarc.nasa.gov/fitsio/fitsio.html}, GSL\footnote{http://www.gnu.org/s/gsl/}, SuperMongo\footnote{http://www.astro.princeton.edu/$\sim$rhl/sm/sm.html}, IRAF).

To avoid these practical difficulties, we have developed the Tool for Automatic Measurement of Equivalent width (\TAME)\footnote{\TAME\ can be downloaded from  http://astro.snu.ac.kr/$\sim$wskang/tame/}, which is written in IDL\footnote{The Interactive Data Language (IDL) is a cross-platform software, providing support for Microsoft Windows\textsuperscript{\textregistered}, Mac OS X, Linux, and Solaris (http://www.exelisvis.com)} and uses a graphical user interface (GUI). 
\TAME\ can be used with any platform OS on which IDL has been installed, and it contains various features required to adjust the environment of \ew\ measurement such as the local continuum and radial velocity of a star. 
Its semi-automatic mode ($hereafter$, the interactive mode) offers more flexible measurement of the \ew\ as similar to DAOSPEC. 
And its fully automatic mode ($hereafter$, the automatic mode) can simultaneously measure the \ews\ for a large set of lines. 
\TAME\ produces a formatted text file containing the \ew\ result which can be used directly in the abundance analysis code MOOG \citep{MOOG}, in addition to a graphical output file with the fitting results of the local continuum and line profile.

In this work, we describe the procedure by which \TAME\ measures the \ews\ of spectral lines and examine the results of \ews\ obtained by \TAME. 
In Sect. 2, we introduce the user interface and input parameters of \TAME.
In Sect. 3, we explain the automatic processes used to measure the \ew\ with \TAME, such as determining the local continuum, searching for blended lines, and fitting the lines with a Gaussian/Voigt profile.
In Sect. 4, we present the comparison of the manual \ew\ measurements obtained using IRAF and those obtained using \TAME\ for the high-resolution spectra of the Sun, whose atmospheric parameters are well known.
We also discuss the difference between the \ew\ estimated by \TAME\ and the input \ew\ for a synthetic spectrum having different spectral resolutions and S/N ratios.  
In Sect. 5, we summarize the advantages of using \TAME\ along with its performance results.

\section{Interface and Input Parameters} 

\begin{figure}[!ht]
    \plotone{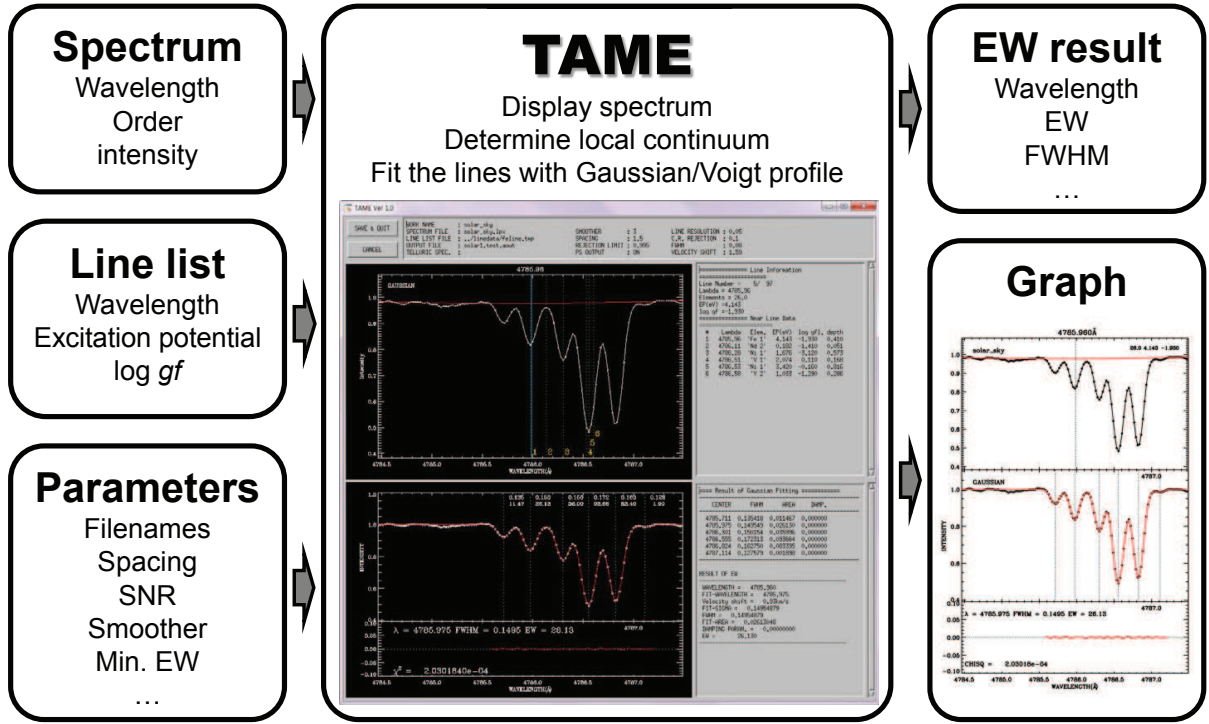}
\caption{Inputs, outputs, and user interface in \TAME\ program. \label{fig:diag}} 
\end{figure}

The inputs, outputs, and user interface of \TAME\ are shown in \figurename~\ref{fig:diag}. 
Initially, \TAME\ requires the spectrum data in text format and the line list file that contains the line information such as the wavelength, element index (for MOOG code; e.g. ''26.0'' for Fe I), excitation potentials (eV), and oscillator strength (log $gf$). 
The other parameters for \ew\ measurement are obtained from the formatted text file, which contains  parameters such as the spacing of wavelength (SPACING), SNR for determining local continuum (SNR), smoothing factor (SMOOTHER), and measurable minimum EW (MINEW).

The SPACING parameter restricts the wavelength window, which is used for determining the local continuum and searching for the lines. 
The SNR parameter is related with the cut-off ratio when the local continuum level is being determined. 
The SMOOTHER parameter is applied to the spectrum in order to mitigate the effect of noise while searching for the lines using numerical derivatives. 
The smoothed spectrum and its derivatives allow the correct lines to be detected in the spectrum even for a low S/N ratio.

\TAME\ is a GUI-based program that provides the plots of the local continuum and the fitting result as shown in \figurename~\ref{fig:diag}.
The GUI consists of the parameter panel ($top$), the interactive panel for adjusting the local continuum ($middle$), and the plotting panel of the line-fitting results and its residuals ($bottom$).
\TAME\ presents the automatic result of the \ew\ measurement that can be interactively adjusted by user. 
Essentially, it has been designed to provide a convenient \ew\ measurement for a large number of absorption lines in a high-resolution spectrum.
It presents the spectrum near the target line and the information of nearby lines adopted from Vienna Atomic Line Database \citep[VALD,][]{kupka00,kupka99,ryabchikova97,piskunov95}. 
After verifying an automatic result, the user can decide whether the \ew\ of the line is reliable. 
If the user prefers, the user can correct the local continuum and adjust the parameters for fitting, such as fitting function (Gaussian/Voigt), smoothing factor, and radial velocity. 
In the automatic mode, \TAME\ can also calculate the \ews\ of all target lines simultaneously, based on the default options described in the formatted parameter file. 

\section{Methods}

\TAME\ measures the \ews\ of spectral lines with following steps: 
\begin{enumerate}
 \item Determine the local continuum near the target line
 \item Normalize the local spectrum with the local continuum 
 \item Identify the lines near the target line in the normalized spectrum 
 \item Fit the target line with the Gaussian/Voigt profile with deblending, if required 
 \item Estimate the \ew, FWHM, and the center wavelength of the target line from fitting.
\end{enumerate}

In this section, we describe the main processes of the \TAME\ program in detail, such as determining the local continuum level, searching for the lines, and fitting the target line with the sample synthetic spectrum, which has a resolving power of R $=$ 10000 and a S/N ratio of 100. 
We selected the resolving power and the S/N ratio of synthetic spectra as the lower limit of which \TAME\ works acceptably.
Then, we briefly discuss the two types of output files generated by \TAME\ and demonstrate how \TAME\ works for actual spectra by using the examples of metal-rich and metal-poor stars.

\subsection{Determining the Local Continuum}

\TAME\ determines the local continuum near the target line according to the following steps: 
\begin{enumerate}
   \item Find the fitting curve in the trimmed spectrum by using the SPACING parameter with a polynomial function (order $=$ 2) 
   \item Cut off the points below $(the~curve) \times \{1 - 2 /\mathrm{(SNR~parameter)} \}$  
   \item Derive a new curve by polynomial fitting with the residual points after the cut-off 
   \item Iterate step 2 $-$ 3 until no points remains that need to be cut off 
\end{enumerate} 

\begin{figure*}[!tb]
    \centering 
    \includegraphics[width=\textwidth]{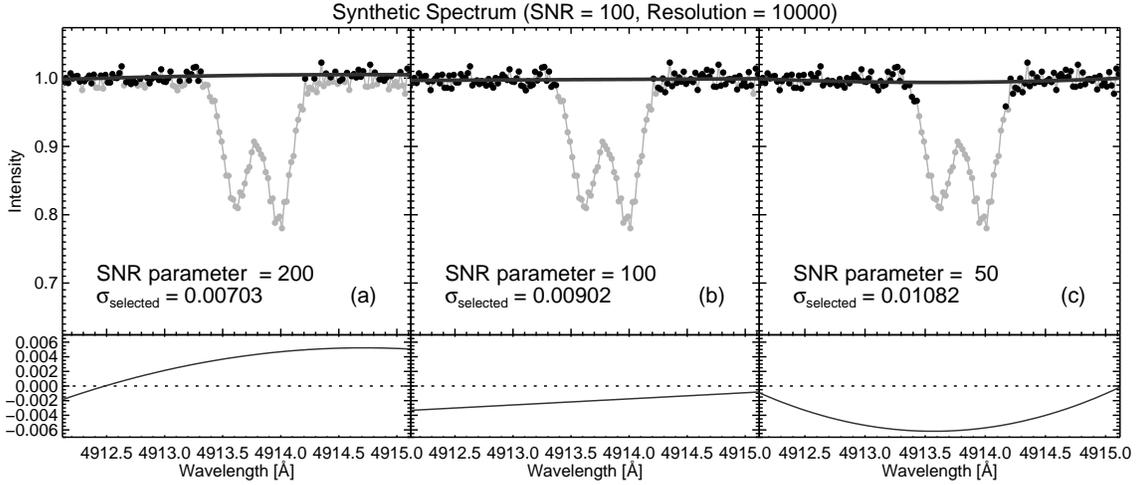}
\caption{The local continuum fitting result for each SNR input parameter ($=$ 200, 100, 50). The $gray$ points show the original synthetic spectrum  with SNR $=$ 100, and the $black$ points represent those  selected by the SNR input parameter for local continuum fitting. The plots at the $bottom$ show the residuals between the original continuum of synthetic spectra ($=$ 1) and the local continuum determined by each SNR parameter. \label{fig:cont_fit}} 
\end{figure*}

We verified the process employed to determine the local continuum by using a synthetic spectrum whose S/N ratio $=$ 100, for different SNR parameter values of 200, 100, and 50 (\figurename~\ref{fig:cont_fit}). 
If the S/N ratio were 100 and the noise followed a normal distribution, the $\sigma_{noise}$ of the continuum in the normalized spectrum would be 0.01 ($ = \sigma_{noise} = 1/\mathrm{SNR} $).
Therefore, we would obtain the final data points representing the local continuum after recursively cutting off the points below the $2\sigma_{noise}$\footnote{If noise follows a normal distribution, after the iterations, points in the upper 50\% (all points above the fitting curve) and lower 47.7\% (the points between the fitting curve and its lower $2\sigma_{noise}$ ) will remain on the final local continuum.} value of the polynomial fitting curve.
Factually, we cannot completely reproduce the original continuum of the spectra only from the observed data. 
It would only be possible to predict the practical local continuum by using S/N ratio of the observed spectrum.
When using the SNR parameter $=$ 100 (\figurename~\ref{fig:cont_fit}b), we found that the final local continuum had a good agreement with the original local continuum, which was supposed to be unity, within a deviation of around 0.3\% ($\sim$0.003). 
The standard deviation of the residual points ($black$) after iterations was very close to 0.01 ($=$ S/N ratio of the synthetic spectrum). 
In the other cases, in which the SNR parameter $=$ 200 and 50 (\figurename~\ref{fig:cont_fit}a and \ref{fig:cont_fit}c), the final local continuum appeared to be more curved than that for SNR $=$ 100. 
In \figurename~\ref{fig:cont_fit}a, the local continuum is determined at 0.5\% higher than the original continuum, and moreover, the shape of the continuum is highly tilted around the target line.
This large discrepancy near the boundary region is because a large number of points were excluded by the input parameter condition, SNR $=$ 200. 
In contrast, as shown in \figurename~\ref{fig:cont_fit}c, when a large number of points were included for continuum fitting, their continuum severely descended around the center. 

The local continuum problem is one of the main causes of \ew\ errors \citep{DAOSPEC}.
Moreover, the errors arising from the local continuum cannot be completely quantified or predicted even though they are known to exist. 
Hence, the best method to determine the local continuum is visually, by experts who are highly disciplined with the stellar spectrum.
However, it would be inefficient and time-consuming to visually examine the local continuum for hundreds of lines only by eyes.

Therefore, \TAME\ enhances the process used to determine the local continuum by using the interactive mode. 
\TAME\ initially suggests the local continuum near the line, which is automatically determined by SNR parameter. 
This local continuum level, which is numerically estimated, can be finely tuned by the user's interaction.
By pressing the ''u'' or ''l'' key, the local continuum level can be shifted up or down proportional to one fifth of $1/(\mathrm{SNR~parameter})$.  
Eccentric points, which are suspected to be produced by the contamination from cosmic rays or bad pixels, can be excluded manually by pressing the ''d'' key for that point. 
When the ''c'' key is pressed, \TAME\ enters the custom mode, which makes it possible to add points anywhere the user wants. 
This adjustment is then directly applied to the fitting result, and hence, it can be quickly verified by the user without the requirement of further operations.  

\subsection{Identifying the Lines}

After determining the local continuum, \TAME\ numerically identifies the absorption lines in the normalized spectrum.
For detecting the center of the line in an arbitrary spectrum, we adopted the method that uses the numerical derivatives and has been suggested by \citet{ARES}.
\TAME\ identifies the line center using the following steps:

\begin{enumerate}
 \item Determine the region for searching for lines, which appears to include all blended lines with the target line
 \item Calculate the 2$^{nd}$ and 3$^{rd}$ derivatives of the normalized spectrum smoothed with  the SMOOTHER parameter 
 \item Find the transition points in the wavelength, where the 3$^{rd}$ derivative changes from positive to negative near the local maximum of the 2$^{nd}$ derivative
 \item Calculate the exact wavelength using the linear interpolation near the transition points   
\end{enumerate}
 
\begin{figure*}[!tb]
    \includegraphics[width=0.32\textwidth]{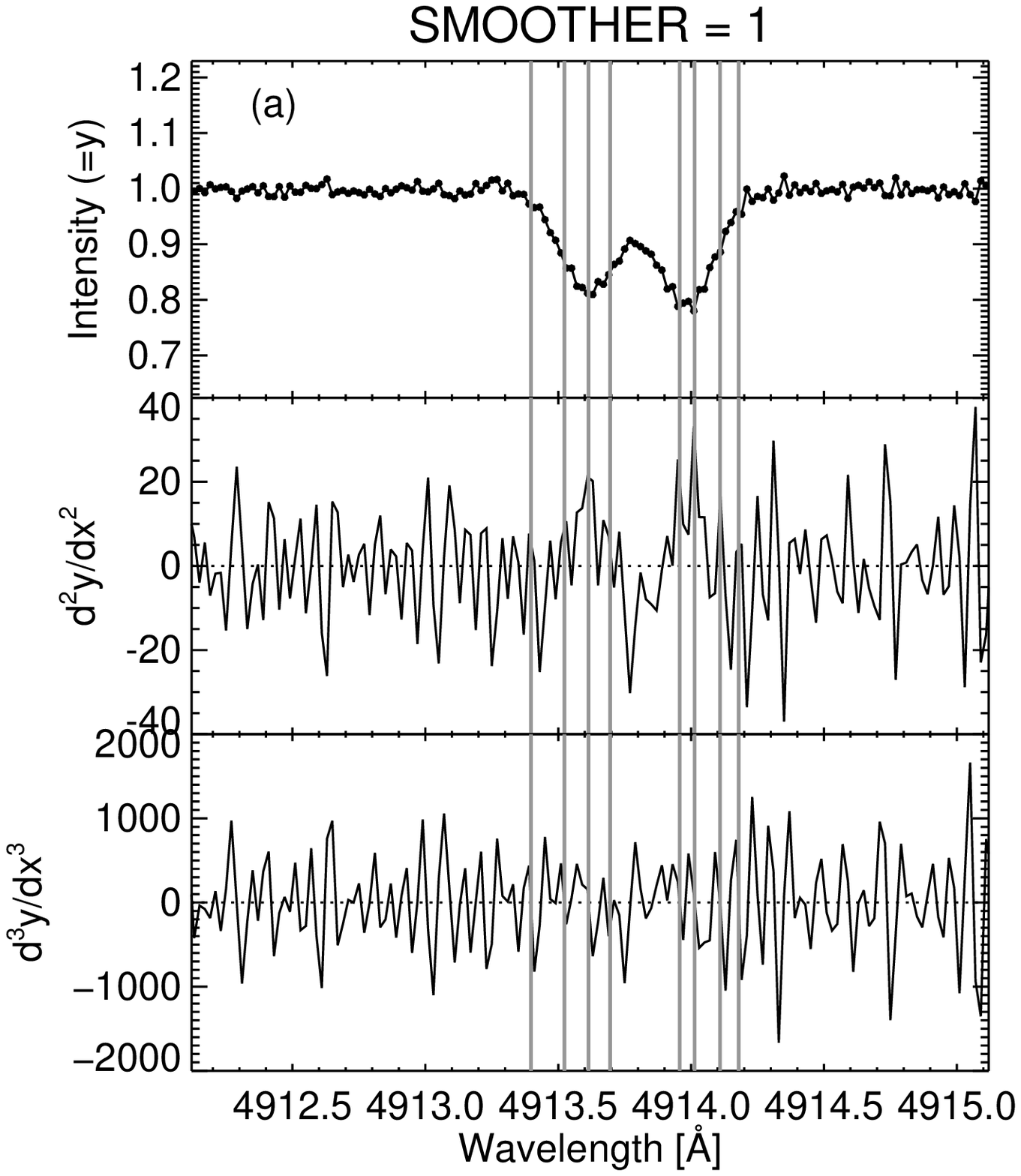}
    \includegraphics[width=0.32\textwidth]{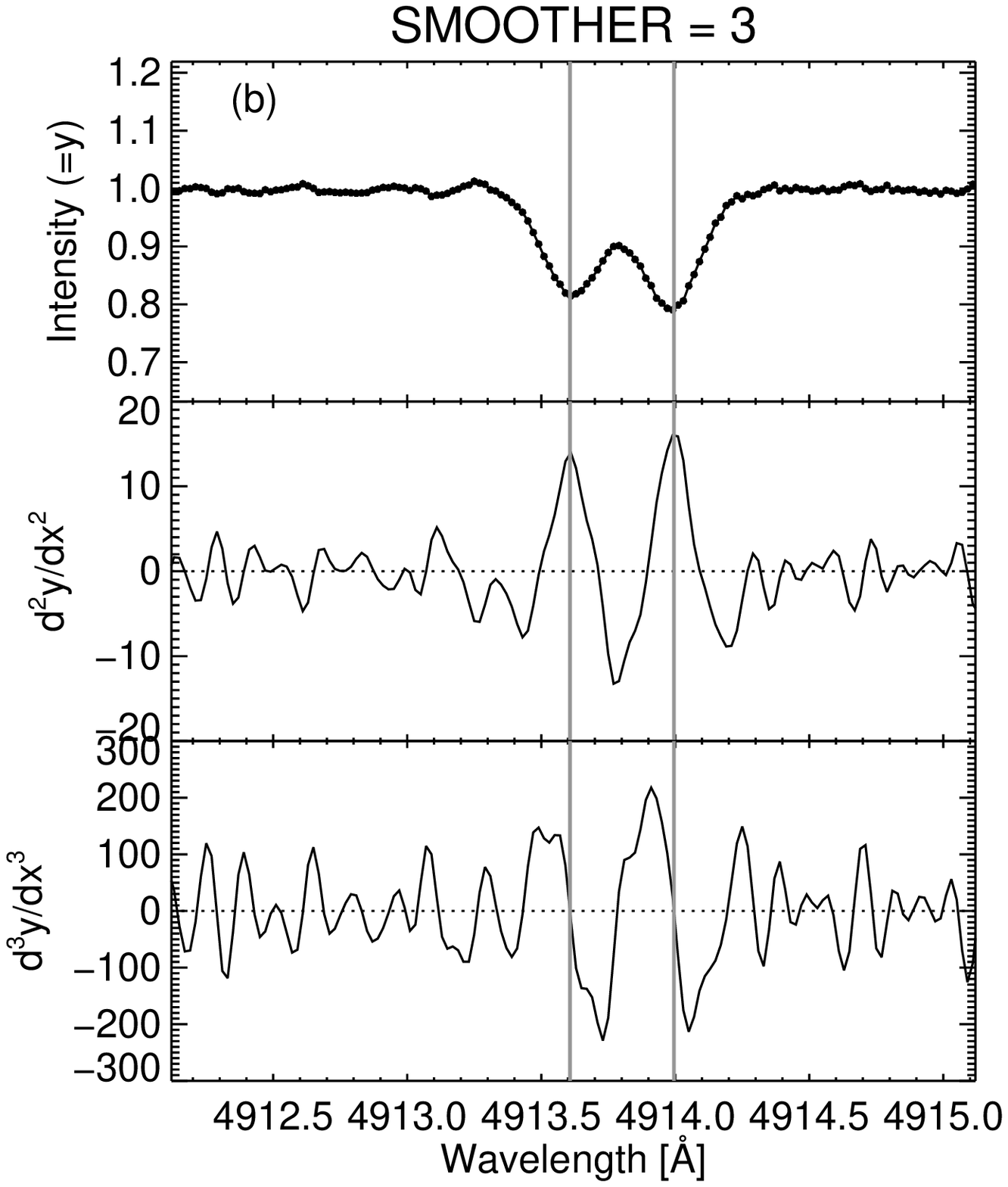}
    \includegraphics[width=0.32\textwidth]{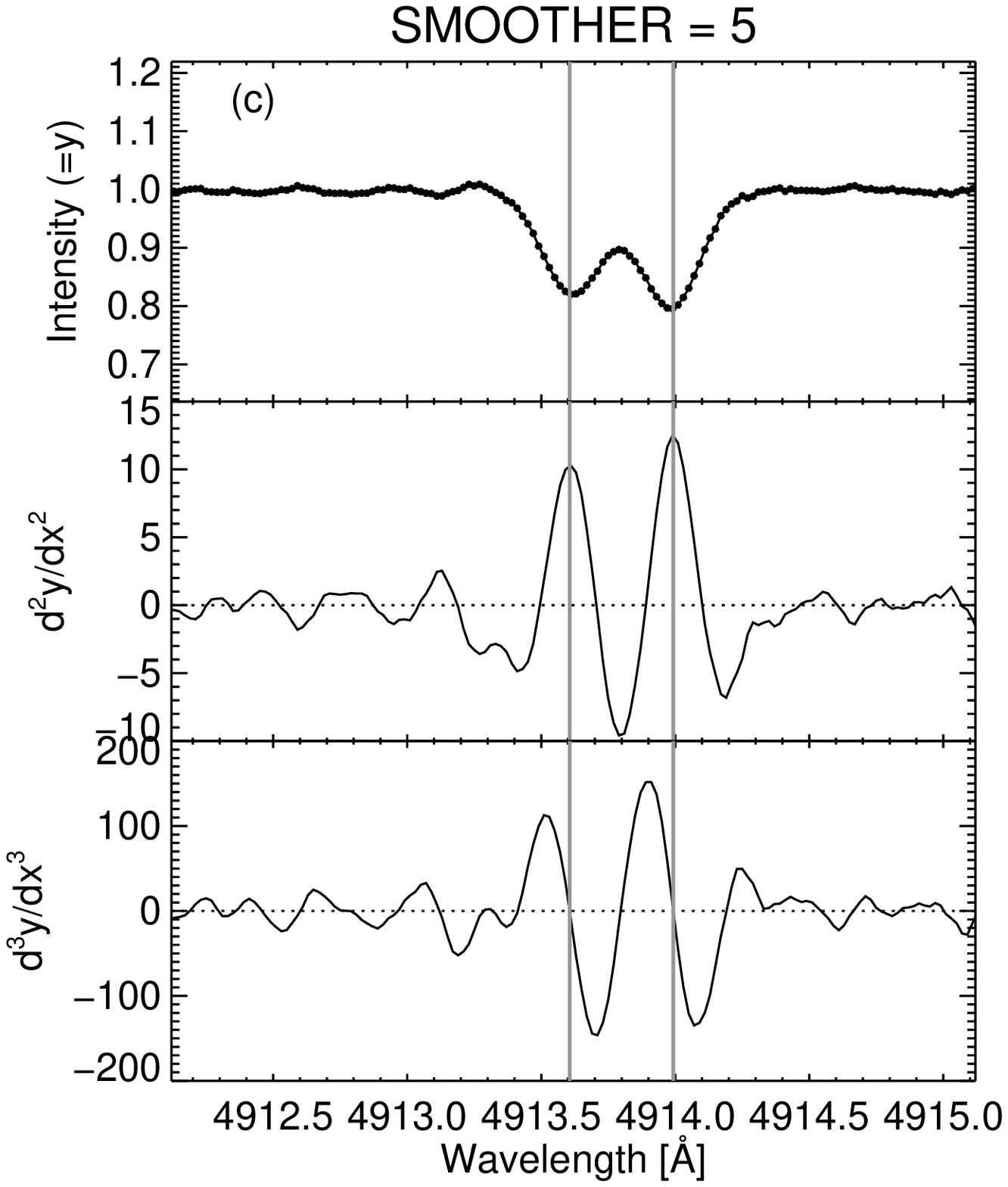}
\caption{The numerical 2$^{nd}$ and 3$^{rd}$ derivatives of the synthetic spectrum for each smoother parameter ($=$ 1, 3, 5). The synthetic spectrum used for the test has a resolving power of 10000 and  a S/N ratio of 100. 
 \label{fig:derivs}} 
\end{figure*}

The numerical derivatives become much more noisy than the original spectrum, because of the noise divergence in a numerical calculation. 
\figurename~\ref{fig:derivs} shows how the noise in the spectrum diverges in the cases without or with smoothing. 
Even when the noise is extremely small in the normalized spectrum, it rapidly multiplies in each  numerical derivative calculation.
In the case of no smoothing (\figurename~\ref{fig:derivs}a), much more lines are identified due to the noisy derivatives, which is amplified in each step of calculation. 
When the spectrum and derivatives were smoothed with 3 or 5 points\footnote{We used a boxcar average for smoothing, $R_i = {1 \over w}\sum_{j=0}^{w} I_{i+j-w/2}$, $w = \mathrm{smoothing~width}$ } (\figurename~\ref{fig:derivs}b and \ref{fig:derivs}c), \TAME\ properly detected two of the correct absorption lines even though the two lines were blended with each other. 
The input wavelengths of these two lines in the synthetic spectrum are 4913.62 and 4913.98 \AA, and \TAME\ finely estimated the wavelengths of these lines at 4913.61 and 4913.99 \AA\ with an error of only 0.01 \AA.

\subsection{Gaussian/Voigt Fitting} 

After searching for absorption lines, \TAME\ calculates the \ew\ of the target line by fitting with a Gaussian/Voigt profile. 
Based on the wavelengths of detected lines through the previous processes, \TAME\ finds the best-fitting of the spectral lines  by least-squares curve fitting with the {\tt mpfit} IDL library\footnote{Markwardt IDL library by Craig Markwardt. Fitting routines are described in the webpage of http://www.physics.wisc.edu/$\sim$craigm/idl/idl.html. For least-squares curve fitting, MPFIT library used the Levenberg-Marquardt algorithm (LMA), originated from MINPACK Fortran library written by Jorge Mor\'e, Burt Garbov, and Ken Hillstrom. }. 
\TAME\ outputs the \ew\ and FWHM of the spectral line as follows: 

\begin{enumerate}
 \item Generate the model functions of spectral lines based on the number of detected lines. For example, if two lines are detected and the user plans to use a Gaussian profile, \TAME\ makes the model functions including two Gaussian profiles.
  \begin{equation}
   \mathcal{F}_{model} = 1 - \sum_{i=1}^2 \mathrm{Gaussian}(\mathrm{wavelength}_i, \mathrm{FWHM}_i, \mathrm{EW}_i)~, 
  \end{equation}
 \item Set the initial value and the range of each parameter such as the wavelength and FWHM. 
 \item Find the best fit by starting with the initial value of each parameter. 
   
\end{enumerate}

\begin{figure}[!tb]
    \centering
    \includegraphics[width=0.7\textwidth]{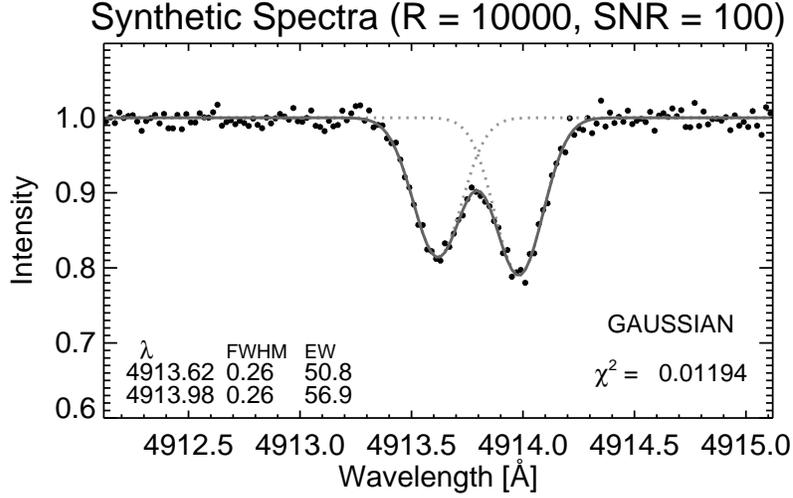}
\caption{The result of Gaussian profile fitting. The $gray$ solid line indicates the best-fitting result for all lines and the $gray$ dotted line represent the fitting profile of each line component.  
 \label{fig:fitting}} 
\end{figure}

\figurename~\ref{fig:fitting} shows the fitting results of the synthetic spectrum for a Gaussian profile. 
As shown in \figurename~\ref{fig:fitting}, \TAME\ estimates the centers of two lines at 4913.62 and 4913.98 \AA\ by Gaussian profile fitting, and obtains the \ews\ of those lines at 50.8 and 56.9 m\AA.  
Considering that the input \ews\ of these lines are 49.8 and 56.2 m\AA, the estimated \ews\ are in very good agreement with the input \ews, having differences within only 1 m\AA. 
We confirmed that despite of the line deblending, \TAME\ can estimate the accurate \ews\ in the synthetic spectrum with R $=$ 10000 and the S/N ratio $=$ 100.

\subsection{Outputs}

After the completion of the \ew\ measurement, \TAME\ outputs a text file and a graphical post-script file. 
The text output file is written in the same format as the input file of the {\tt abfind} driver of MOOG code \citep{MOOG}, in order that the user can instantly calculate the abundances of lines using MOOG code. 
This text output also contains the line center and FWHM of the fitting profile, and the radial velocity calculated by the difference between the rest-frame wavelength and the measured wavelength for each line. 
The representative radial velocity of the star can be derived by averaging the radial velocities of individual lines along with their standard deviation. 
Additionally, the FWHM of the line and the $\chi^2$ value from the fitting could be used for the diagnostics of lines. 
Each atomic spectral line has a specific FWHM for each wavelength, atomic mass, and stellar effective temperature. 
Extremely broad or extremely narrow lines, relative to the others, can be expected to be affected by obscure lines or to not originate from the stellar atmosphere, and therefore, they can be neglected in the \ew\ results. 
This is because the broad lines, that are closly blended each other, cannot be deblended due to the limit of the spectral resolution.

The graphical output file shows the spectrum and fitting plots of the local continuum and the line profile for each line.
Because \TAME\ generates these plots in a post-script file, the user can easily open and print this graphical output. 
When performing the abundance analysis, the graphical output is very useful for checking whether the abundance of each line is well determined.
The \ews\ of more than tens of lines are generally adopted for the chemical abundance of one element. 
If the abundance of a line is located far from the abundance distribution of the other lines of the same element, the graphical output can help with inspecting the spectrum around aberrant lines and their fitting plots.

\subsection{Examples using the Spectra of Metal-rich and Metal-poor Stars} 

We examined the process of \ew\ measurement performed by \TAME\ with the actual spectra of HD 75732 ([Fe/H] = +0.35) as a metal-rich star and HD 155358 ([Fe/H] = -0.63) as a relatively metal-poor star. 
These spectra were obtained with the BOAO Echelle Spectrograph (BOES) \citep{kim02,kim07} in 2008 and have the spectral resolution R $=$ 30000 and an S/N ratio of $\sim$200 \citep{kang11}. 

\begin{figure*}[!tb]
    \centering
    \includegraphics[width=0.48\textwidth]{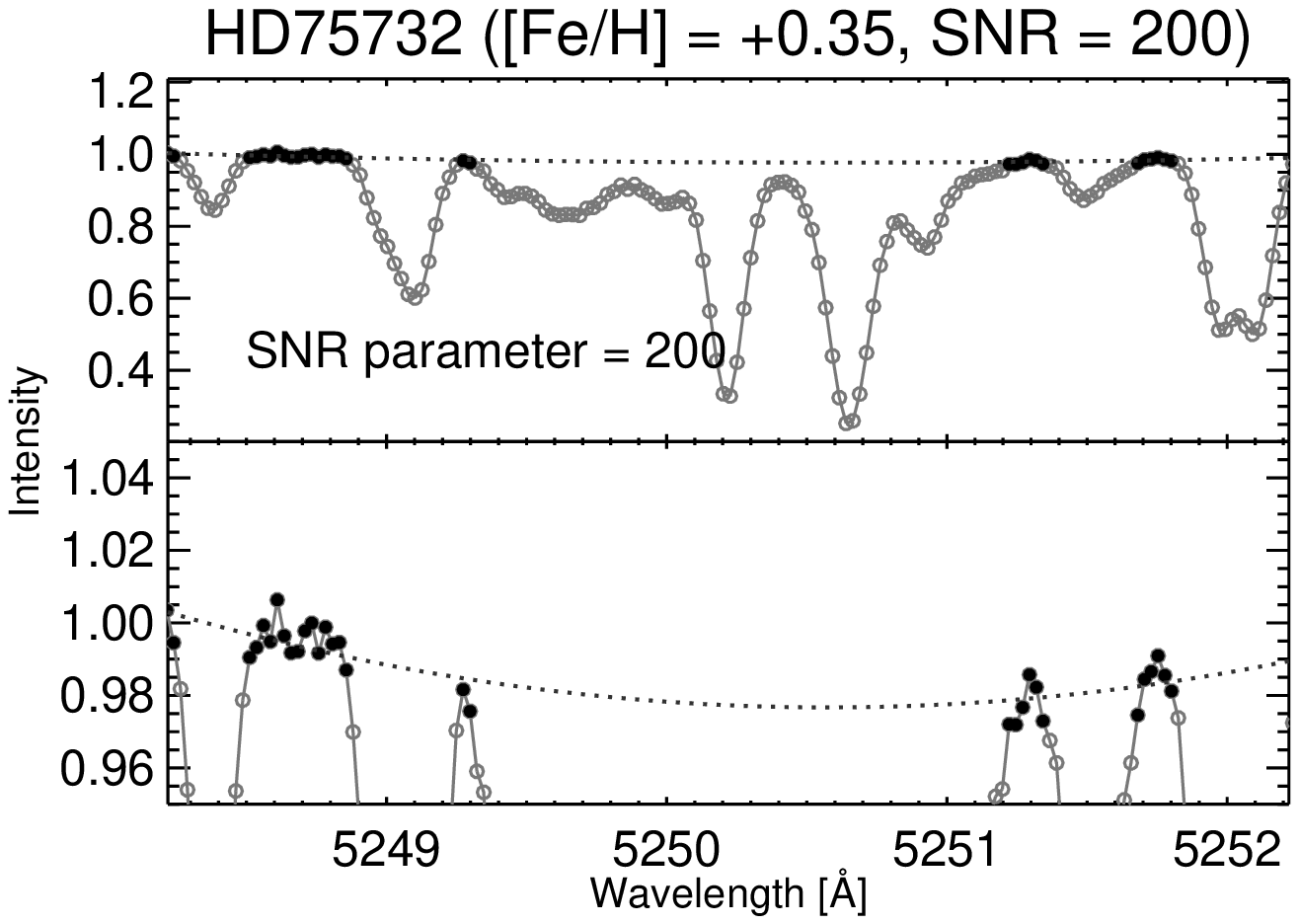}
    \includegraphics[width=0.48\textwidth]{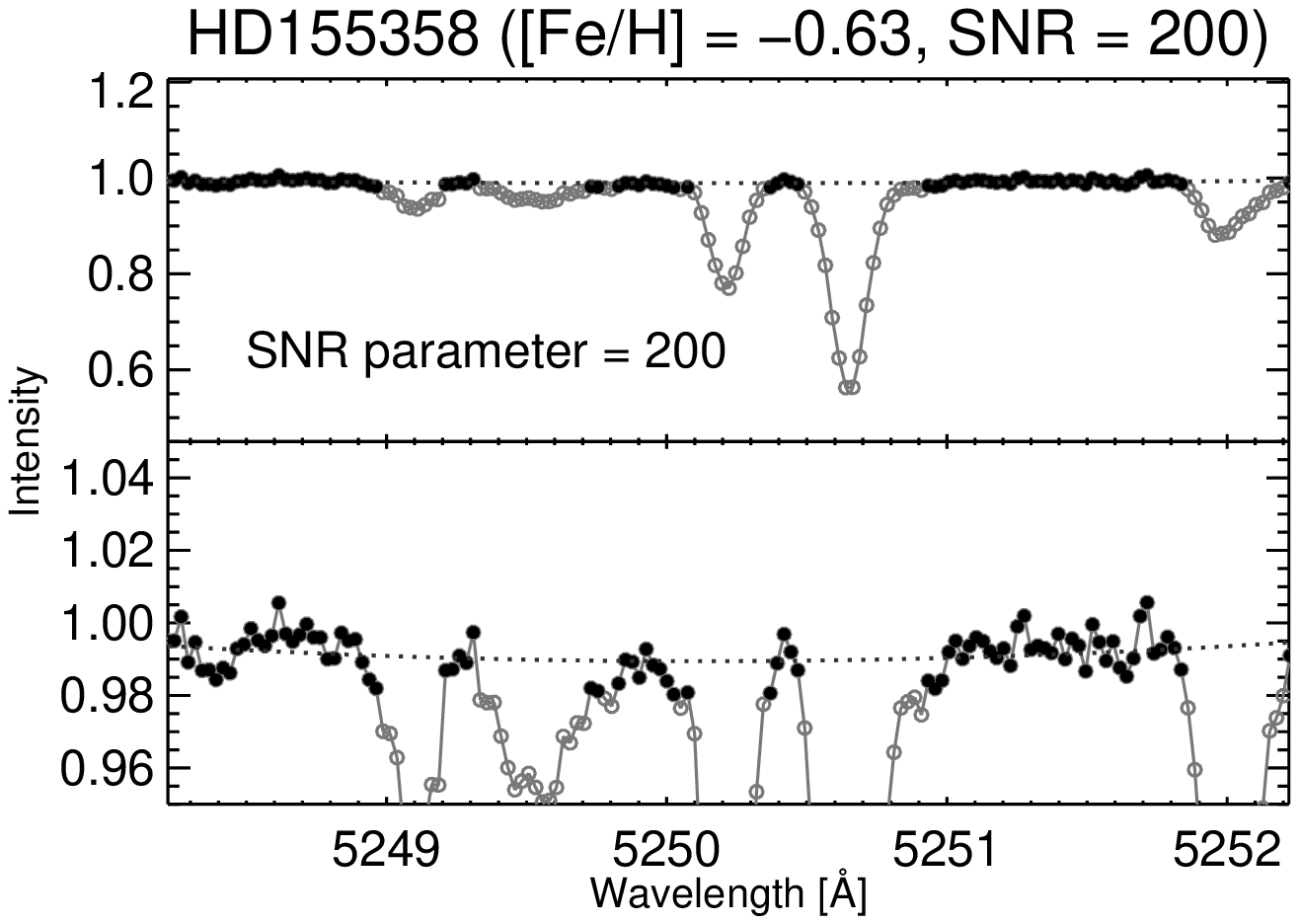}
\caption{The determination of the local continuum for the two cases of a metal-rich and a metal-poor star, i.e., HD 75732 and HD 155358. The  $lower$ panels show the magnifying results around the continuum.
 \label{fig:ex1}} 
\end{figure*}

\figurename~\ref{fig:ex1} shows the local continuum fitting results for the \FeI\ line at 5250.22 \AA\ for two sample stars.
For the metal-rich star HD 75732, the central region of the local continuum appears to sink below about 2\%, relative to the side region.
This is, as mentioned above, because of the undersampling of the points that are used to determine the local continuum in this spectral region. 
It is difficult to manage this bending local continuum in a crowded region with computational manipulation, because neglecting a very large number of points by a statistical method might produce an unstable fitting result.
The best solution for metal-rich stars such as HD 75732, is to confirm the fitting results visually and to correct them through manual interaction. 
Further, we observed that the local continuum of the metal-poor star HD 155358 was well determined.

\begin{figure*}[!tb]
    \centering
    \includegraphics[width=0.48\textwidth]{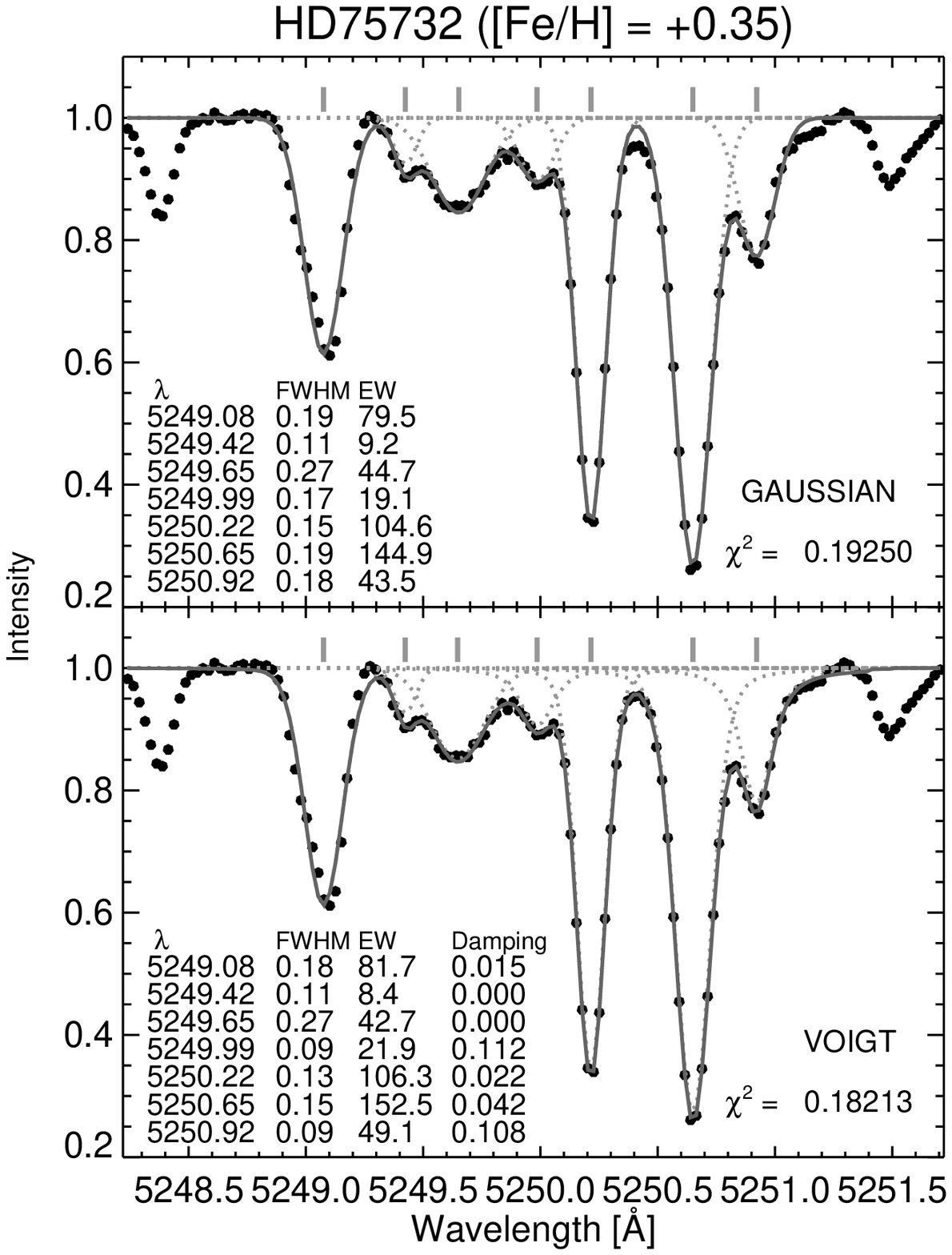}
    \includegraphics[width=0.48\textwidth]{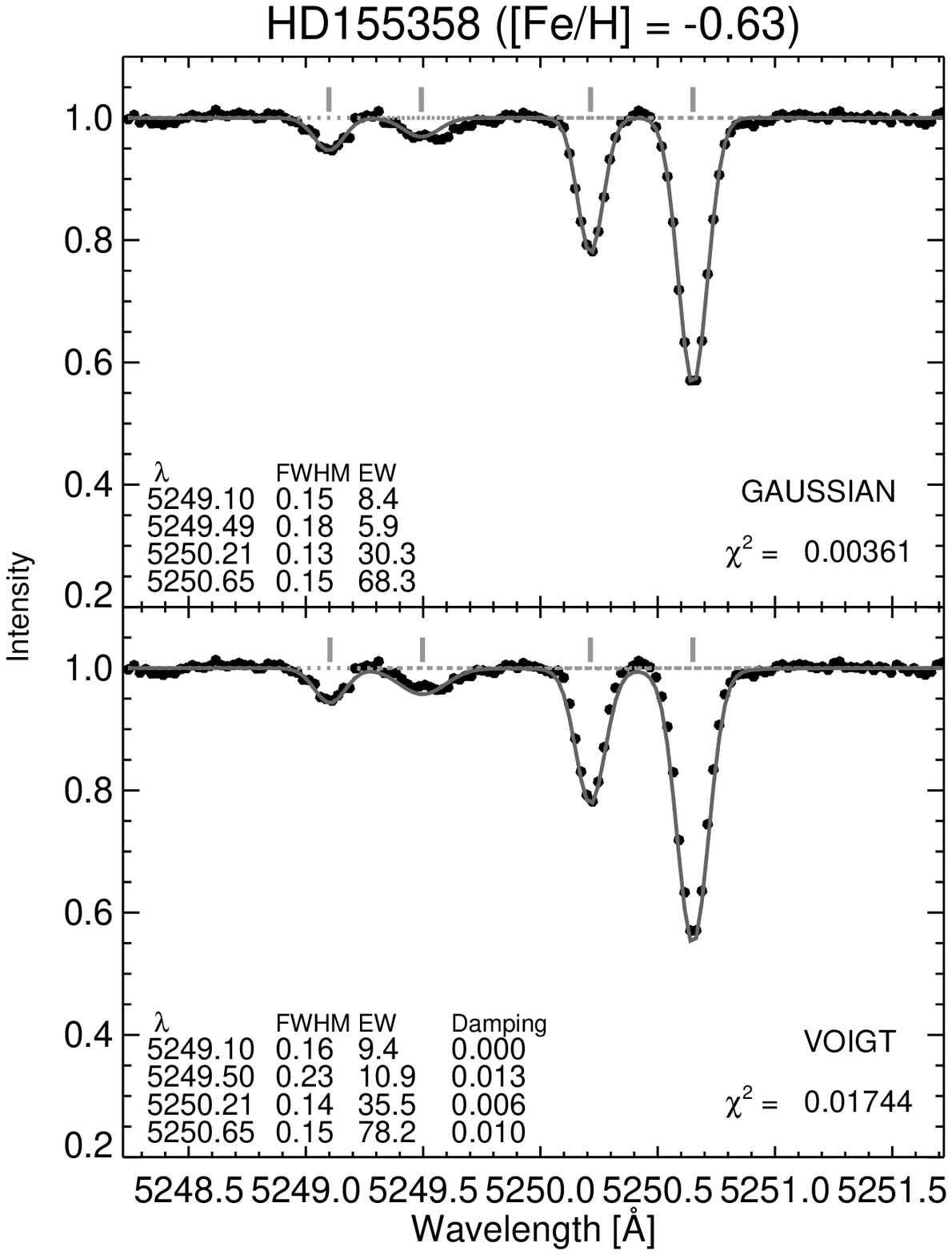}
\caption{The line detection and fitting results of HD 75732 and HD 155358. The $black$ dots represent the observed spectrum, and the $gray$ vertical thick lines indicate the positions of detected lines. The $gray$ dotted lines and $gray$ solid lines denote the fitting profile for each line and for all the lines, respectively. 
 \label{fig:ex2}} 
\end{figure*}

\figurename~\ref{fig:ex2} illustrates the fitting results with Gaussian and Voigt profiles. 
For the test, we plotted the fitting results of all the lines near the target line beyond the region that appeared to be blended with nearby lines. 
In normal cases, \TAME\ automatically defines the blended region by comparing the line features with the local continuum level, and performs fitting only with the points in that region.  
In the case of HD 75732, it can be seen that there are several lines that do not match the Gaussian profile. 
The line at 5249.08 \AA\ seems to be blended with a weak line, which cannot be detected by the method that uses numerical derivatives. 
The \ews\ of two lines at 5250.22 \AA\ and 5250.65 \AA\ are more than 100 m\AA, and hence, show a better fitting result when a Voigt profile is used. 
On the contrary, the fitting result of the metal-poor star HD 155358 shows a better $\chi^2$ and fitting result for a Gaussian profile, because the lines for this star are much weaker than those of metal-rich HD 75732. 

As a result, it can be concluded that \TAME\ appears to work acceptably for metal-poor stars, even in the automatic mode, and that careful adjustments in the interactive mode might be required for metal-rich stars or for strong lines. 

\section{Results}

The \ew\ result of \TAME\ has been validated by comparing it with the \ews\ measured by IRAF {\tt splot} task, and by applying \TAME\ to various synthetic spectra. 
The \ews\ in the solar spectrum have been manually measured by IRAF {\tt splot} task and automatically estimated using \TAME. 
Then, we observed whether \TAME\ is appropriate for the abundance analysis by investigating not only  the difference in the \ews\ but also the atmospheric parameters derived with those \ews. 
In order to evaluate the reliability for a variety of spectra, we examined the \ews\ measured by \TAME\ for synthetic spectra having different spectral resolutions and S/N ratios.  

\subsection{Comparison with Manual Measurements}

We estimated the solar \ews\ of atomic lines using \TAME\ and IRAF {\tt splot} task.
The high-resolution spectrum of the Sun has been obtained with BOAO Echelle Spectrograph (BOES) and has spectral resolution of R $=$ 30000 and S/N ratio $\gtrsim$ 300.

\begin{figure}[!tb]
    \centering
    \includegraphics[width=0.5\textwidth]{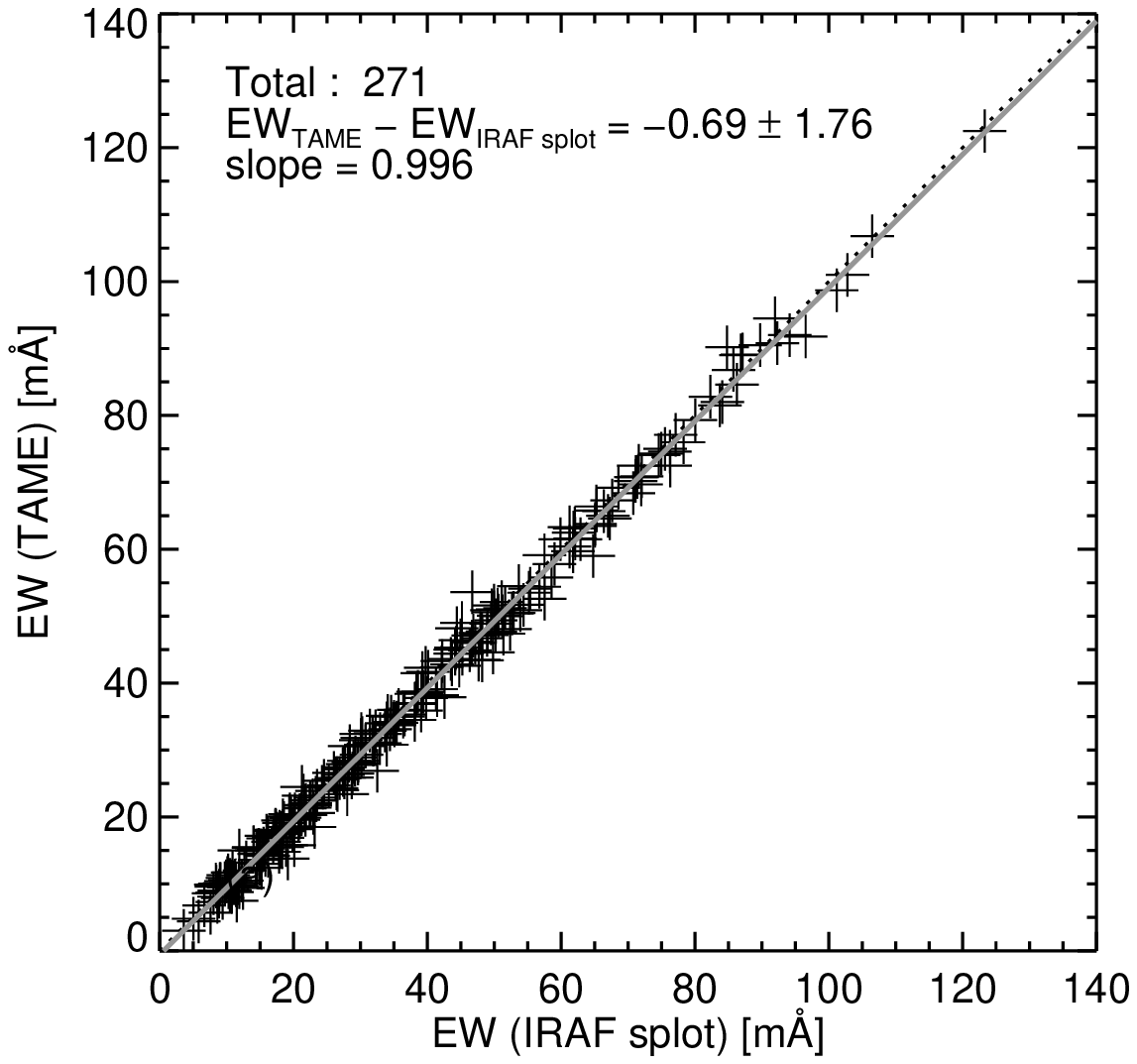}\\
    \includegraphics[width=0.5\textwidth]{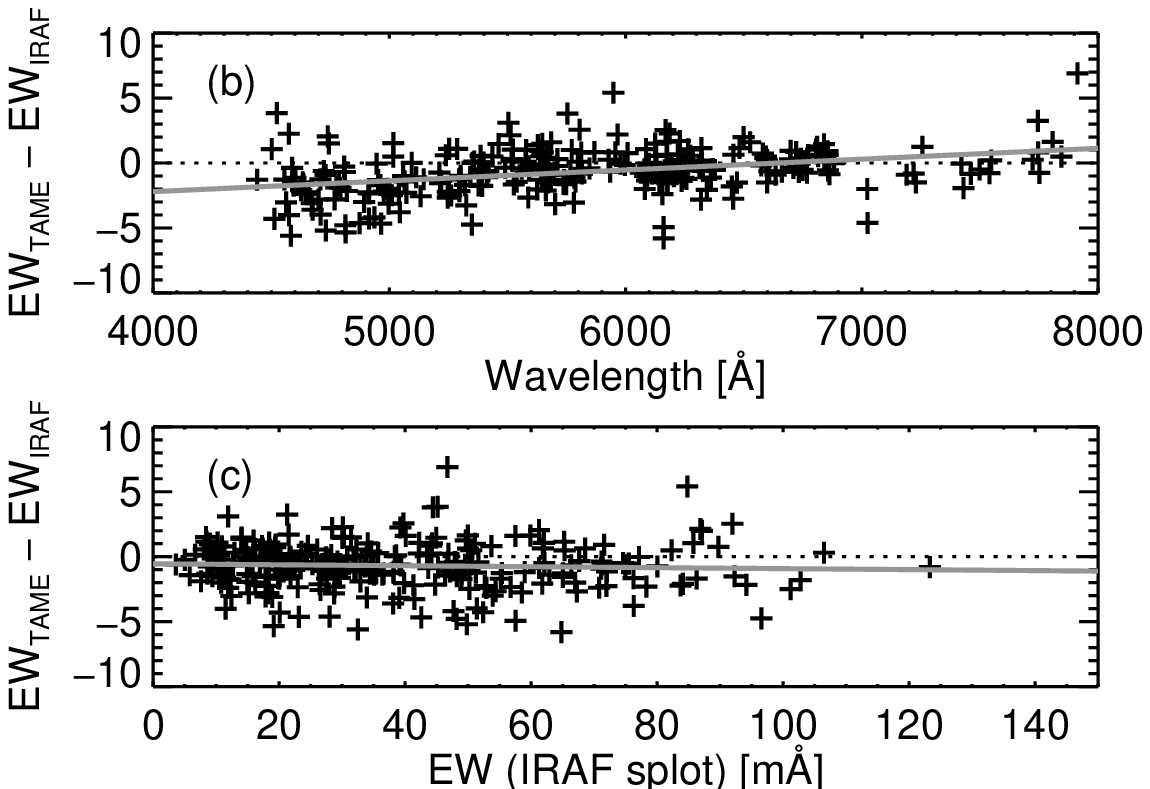}
\caption{The comparison between the \ews\ measured by IRAF {\tt splot} task and those obtained using \TAME. The $top$ panel shows the difference between the two measurements, and the $middle$ and $bottom$ panels show the trends in the \ew\ differences along the wavelength and \ew\ for each line. 
 \label{fig:splot}} 
\end{figure}

\figurename~\ref{fig:splot} shows the result of comparing the \ews\ obtained using IRAF {\tt splot} task and those estimated by \TAME. 
The average and standard deviation between two \ew\ results are acceptable at -0.69 and 1.76 m\AA, respectively. 
In the plots of the \ew\ difference vs. the wavelength and \ew, it appears that the \ew\ differences depend on the wavelength and have no dependence on the \ew. 
At a short wavelength ($\lesssim$ 5500 \AA), as shown in \figurename~\ref{fig:splot}b, the difference in the \ew\ measurement decreases to -5 m\AA\ and becomes more scattered. 
This is largely owing to the local continuum determination in the crowded region, where \TAME\ cannot avoid the undersampling of fitting points.
The undersampling of valid points reduces the local continuum level and can cause the underestimation of  \ews.  
In the comparison between \TAME\ and IRAF {\tt splot} task, we did not manipulate the automatic measurement in order to validate the automatic process used in \TAME. 
The underestimated \ews\ in a crowded region of spectra could be rectified in the interactive mode, by adjustment of the local continuum level. 

\begin{figure*}[!tb]
    \includegraphics[width=0.48\textwidth]{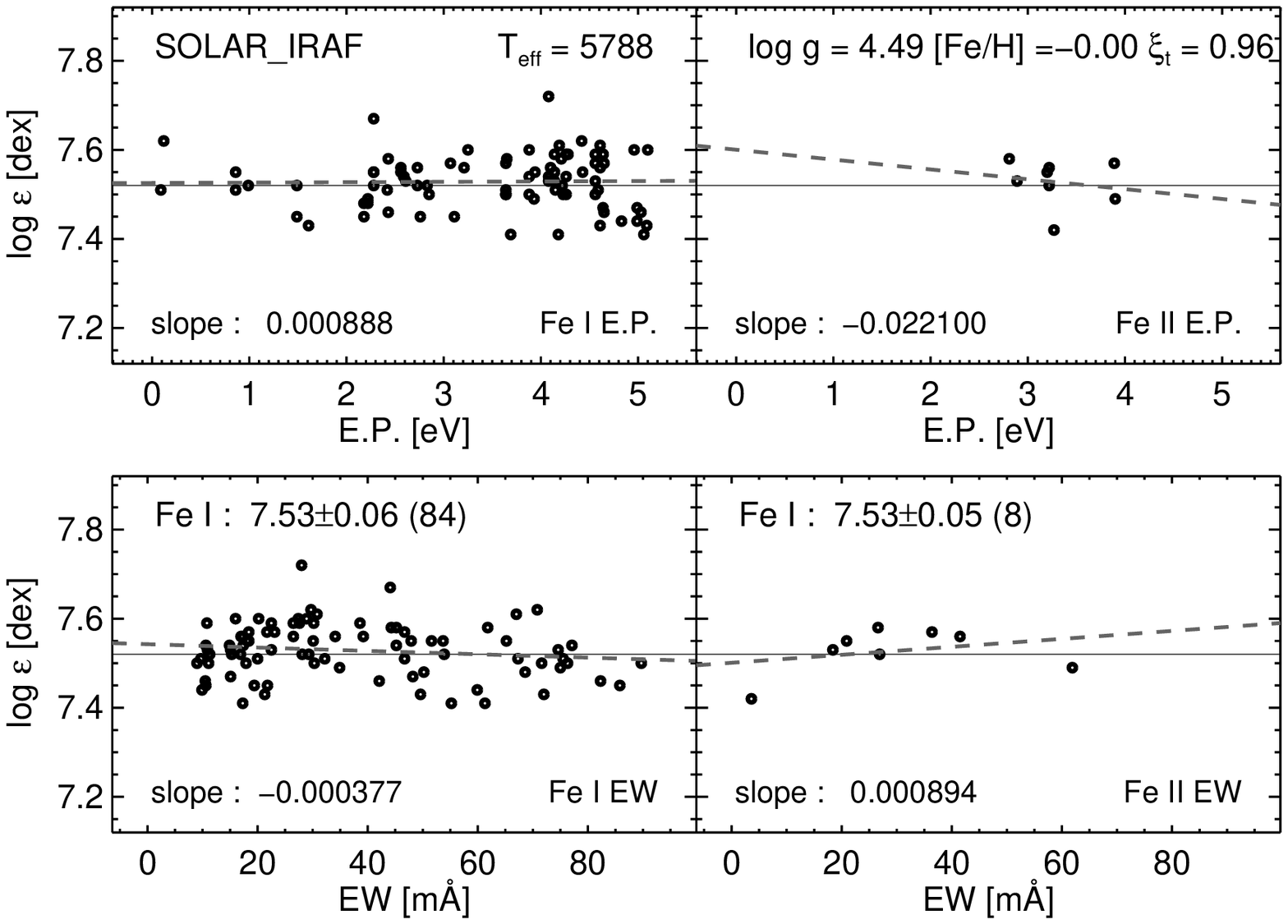}
    \includegraphics[width=0.48\textwidth]{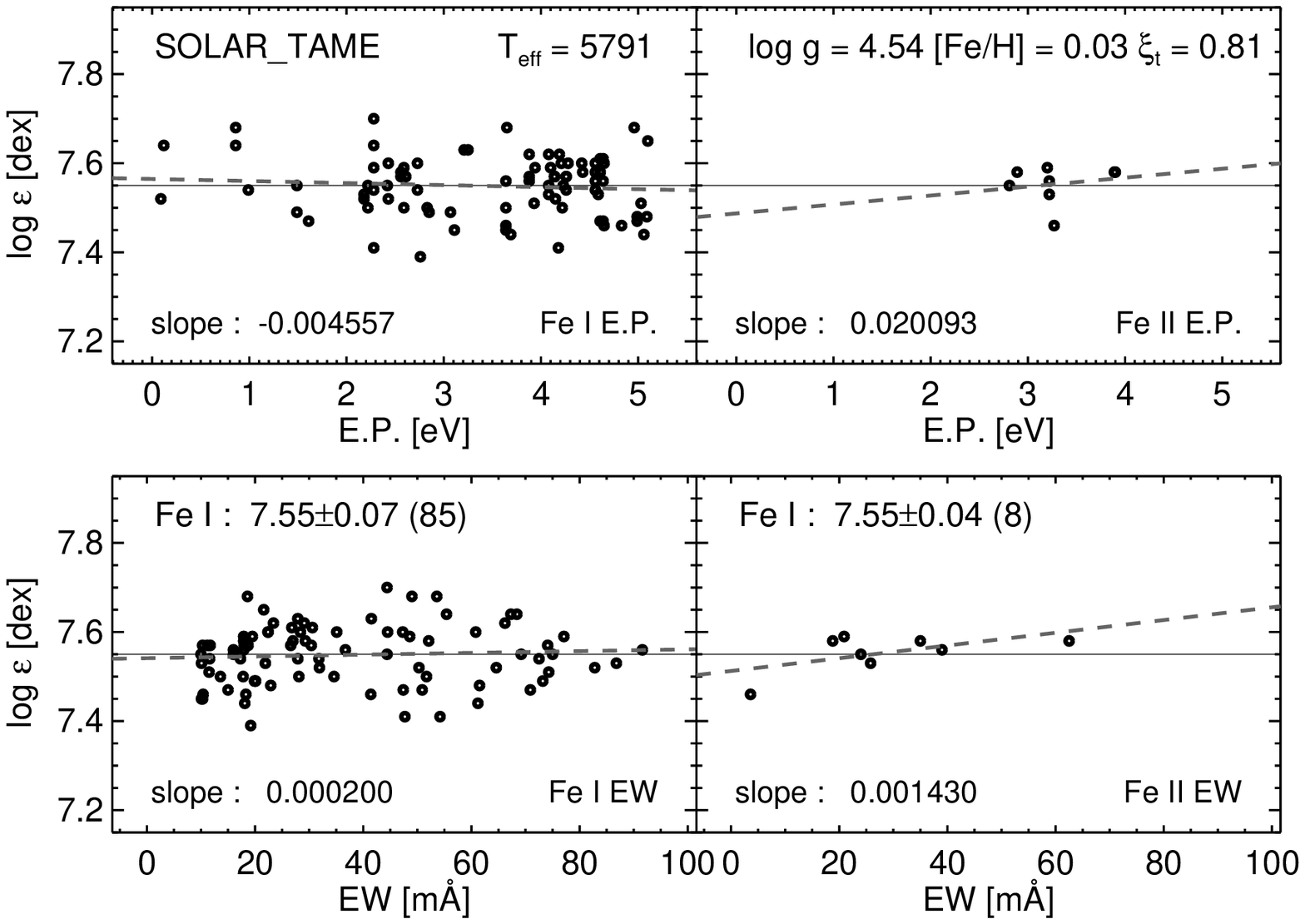}
\caption{The results of the fine analysis performed using the \ews\ of the \FeI\ and \FeII\ lines. The four plots on the $left$ side show the abundances obtained from the \ews\ measured by IRAF {\tt splot} task and the plots on the $right$ show those obtained using the \ews\ measured by \TAME. The dashed lines denotes the linear fitting results.
 \label{fig:param}} 
\end{figure*}

We derived the atmospheric parameters of the Sun from the two sets of \ews\ obtained by IRAF and \TAME.
Using the \FeI\ and \FeII\ lines, we performed the fine analysis, which employs the dependence of  abundance on the excitation potential and \ew\ of each line and the abundance difference between neutral and singly ionized lines. 
We adopted MOOG code and Kurucz ATLAS9 model grids \citep{ATLAS9, castelli04} for estimating the  abundance of each iron line.  
The result of the fine analysis is shown in \figurename~\ref{fig:param}. 
We obtained the atmospheric parameters of the Sun as \teff\ $=$ 5788 K, \logg\ $=$  4.49 dex, [Fe/H] = 0.00 dex, and $\xi_t = $ 0.96 \kms\  from the \ews\ measured by IRAF. 
Those derived from the \ews\ obtained using \TAME\ are \teff\ $=$ 5791 K, \logg\ $=$ 4.54 dex, [Fe/H] = 0.03 dex, and $\xi_t = $ 0.81 \kms.  
The atmospheric parameters derived from the \ews\ measured by \TAME\ are in a good agreement with those derived from the \ews\ measured manually using IRAF {\tt splot} task.

\subsection{Test with Synthetic Spectrum} 

\begin{figure*}[!tb]
    \includegraphics[width=0.32\textwidth]{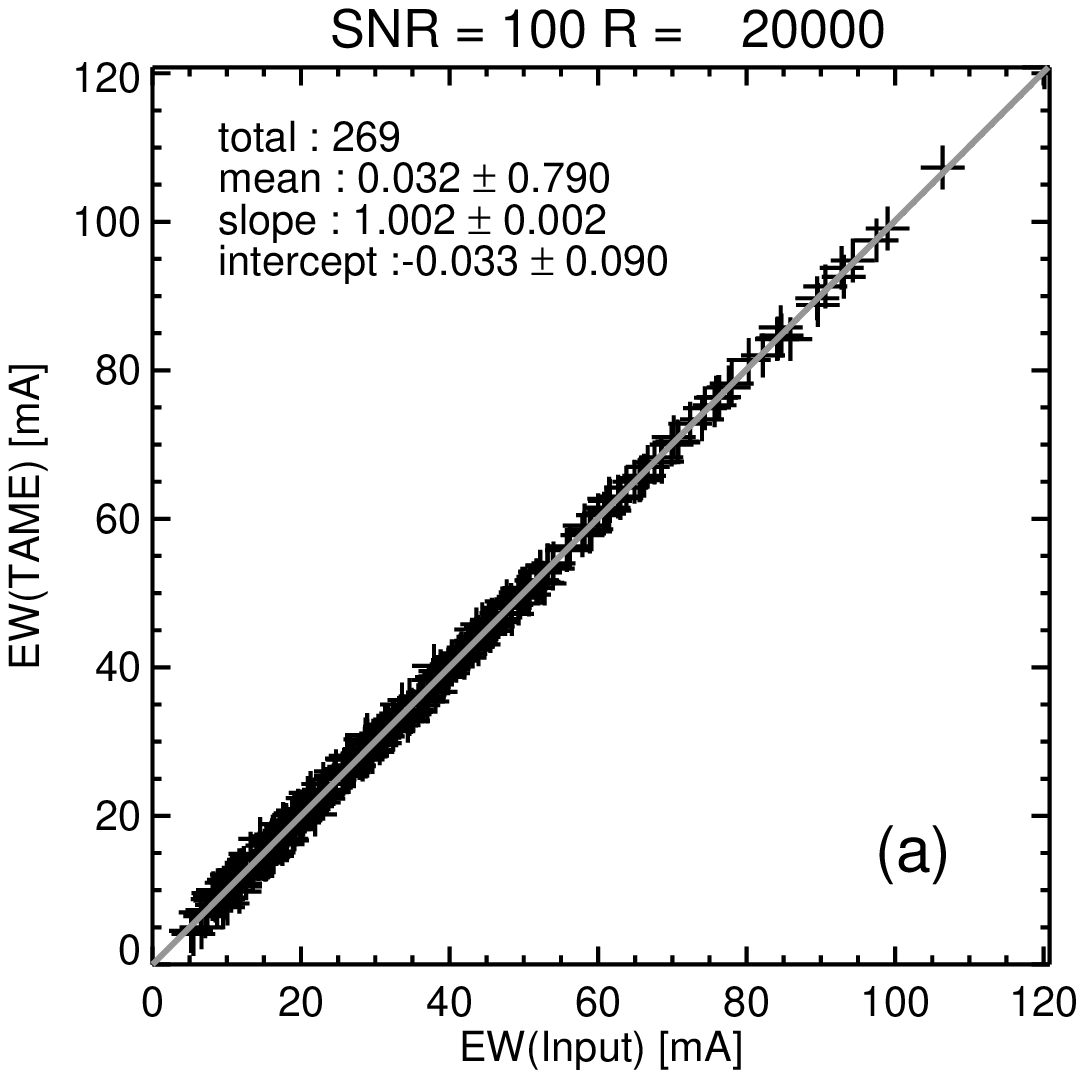}
    \includegraphics[width=0.32\textwidth]{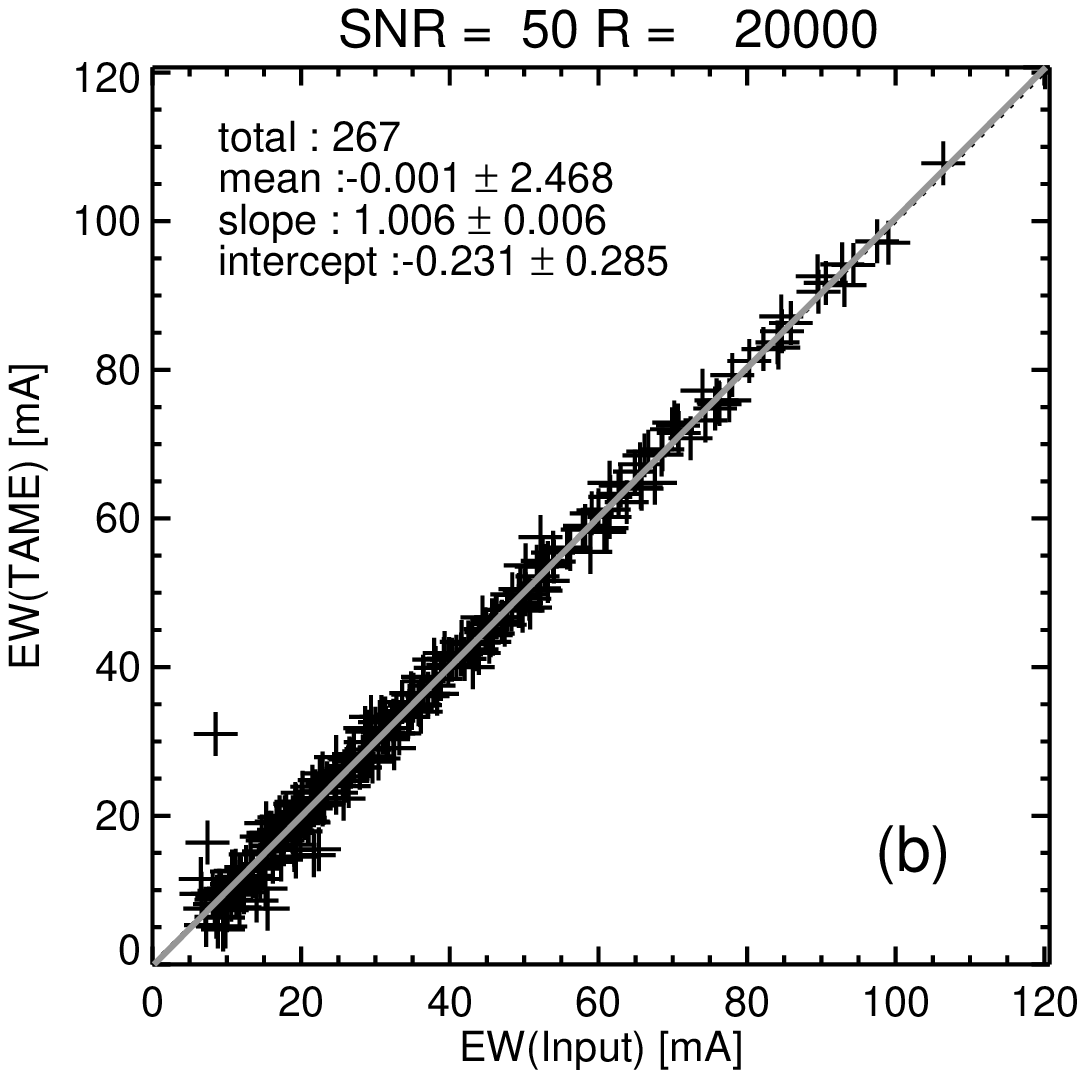}
    \includegraphics[width=0.32\textwidth]{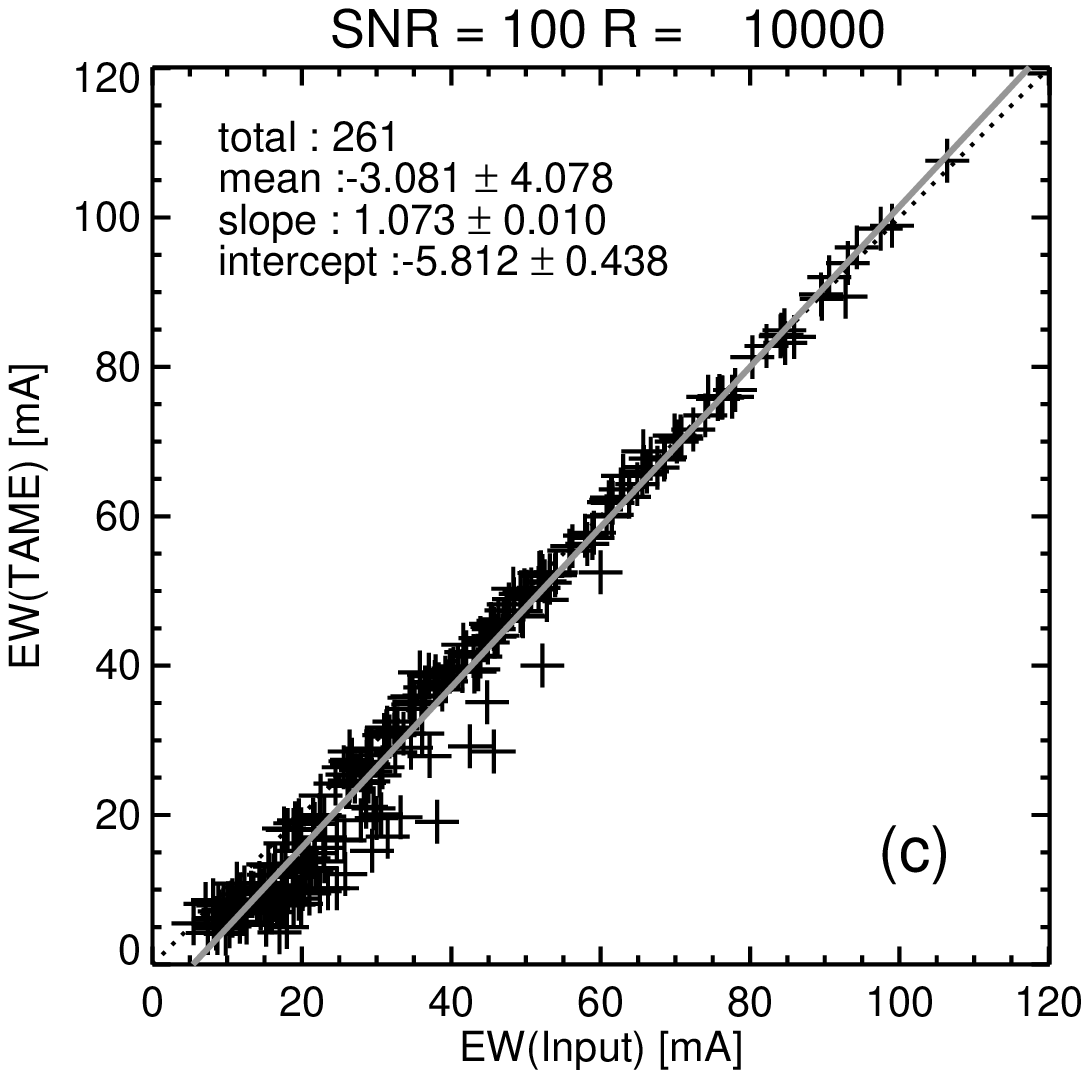}
\caption{The sample plots of the input \ews\ for the synthetic spectra and those measured by \TAME. The plots show the results for three different kinds of synthetic spectra, which have (SNR, R) $=$ (100, 20000), (50, 20000), and (100, 10000) in the case that the SMOOTHER parameter $=$ 3.
The plot in the $middle$ represents the example of a low S/N ratio, and the plot in the $right$ shows how the \ew\ measurement changes when spectral resolution decreases with respect to the plot on the $left$.   
 \label{fig:diff}} 
\end{figure*}

\begin{figure*}[!tb]
    \includegraphics[width=0.95\textwidth]{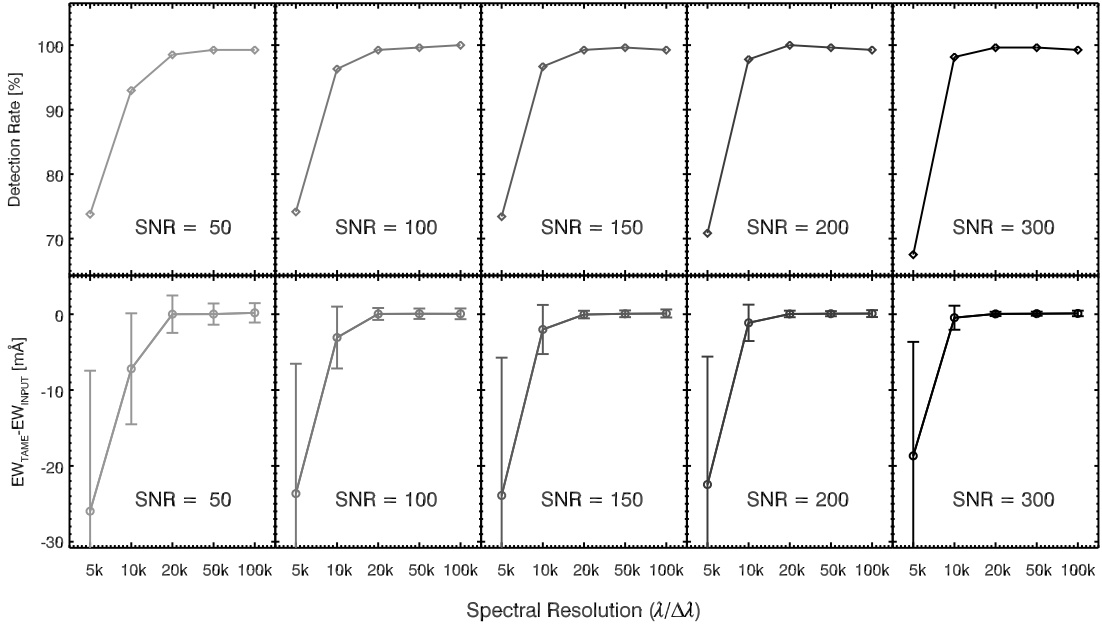}
\caption{The detection ratio of lines and the difference between the input and the measured \ew\ by \TAME\ (for SMOOTHER $=$ 3).  
The error bars in the $lower$ plots represent the standard deviation of \ew\ differences for each spectral resolution and each S/N ratio. 
 \label{fig:trend}} 
\end{figure*}

\ew\ measurement is obviously sensitive to the quality of the spectrum, or in other words, the errors in \ew\ measurement depend on the S/N ratio as well as the spectral resolution of the spectrum. 
Therefore, we considered the result of \ews\ measured by \TAME\ for synthetic spectra with different S/N ratios and spectral resolutions.
For assessing the reliability of \ew\ measurement on the different spectra, we used the synthetic spectra with S/N ratios of 50, 100, 150, 200, and 300, and spectral resolutions R $=$ 5000, 10000, 20000, 50000, and 100000. 
We then measured the \ews\ of the lines for each synthetic spectrum in the automatic mode; that is, we obtained the \ews\ using \TAME\ without any interactive operations.  
\figurename~\ref{fig:diff} shows three examples of input and measured \ews. 
Compared with \figurename~\ref{fig:diff}a, \figurename~\ref{fig:diff}b indicates that the rms of the \ew\ difference increases to $\sim$2.5 m\AA\ with a S/N ratio decreasing from 100 to 50, for the same resolution, R $=$ 20000. 
When spectral resolution decreases from 20000 to 10000, as shown in \figurename~\ref{fig:diff}c, the \ews\ of weak lines appear to be underestimated by \TAME. The underestimation of \ew\ at a low spectral resolution will be discussed further together with \figurename~\ref{fig:trend}. 

In order to investigate the variation in the \ew\ measurement resulting from the properties of the spectrum, we calculated the detection rate of lines and the difference between input and measured \ews\ for each S/N ratio and each spectral resolution. 
As shown in \figurename~\ref{fig:trend}, we found that the \ew\ difference between the input and measured values converged rapidly to zero when the spectral resolution increased (especially, R$\gtrsim$ 20000) and less sensitive to the noise in the spectrum. 
Similarly, as the amount of the \ew\ difference increased, the detection rate also decreased with a decreasing spectral resolution.   
The average \ew\ difference sharply decreased below -15 m\AA\ at low resolution of R $=$ 5000, and particularly, in the case of S/N ratio $=$ 50, the \ew\ difference reduced to -25 m\AA. 
This is because \TAME\ is likely to deblend the target line into arbitrary several lines in the spectrum with a low resolution.  
The depth of the absorption line becomes more shallow when a spectral resolution decreases, and hence the noise patterns can be confused with a feature of absorption line in numerical line identification.
This also explains why the detection rate decreases at a low spectral resolution.  
However, excessive deblending by \TAME\ can be reduced by using a large SMOOTHER parameter in most cases. 
From the data in \tablename~\ref{tbl:stats}, we could confirm that using a higher SMOOTHER parameter ($=$ 5) causes the detection rate to become much higher and the average difference further stabilizes even for a low resolution of R $=$ 5000.

From this assessment, we conclude that \TAME\ is reliable for measuring the \ews\ in the spectrum of a typical high-resolution echelle spectrograph, which has a spectral resolution of R $\gtrsim$ 20000, and the rms result indicates that the error in the \ew\ measurement reduces to less than 1 m\AA\ for a S/N ratio $\gtrsim$ 100.

\begin{deluxetable}{rr rrr c rrr c}
\tablecaption{Line detection ratio and statistics of the difference between the input and measured \ews\ for different spectral resolution and S/N ratio values of synthetic spectra \label{tbl:stats}}
\tabletypesize{\scriptsize}
\tablewidth{0cm}
\tablehead{
 & & 
   \multicolumn{3}{c}{SMOOTHER = 3} & &
   \multicolumn{3}{c}{SMOOTHER = 5} & \\
 \cline{3-5} \cline{7-9} 
  Resolution & SNR  & 
   Detection[\%] & Avg. [m\AA] & rms [m\AA] & & 
   Detection[\%] & Avg. [m\AA] & rms [m\AA] & 
}
\startdata 
      5000 &        50 &     73.80 &   -25.978 &    18.528 & &     68.63 &   -16.308 &    12.448 & \\
      5000 &       100 &     74.17 &   -23.656 &    17.121 & &     86.35 &   -10.704 &    10.558 & \\
      5000 &       150 &     73.43 &   -23.907 &    18.177 & &     88.19 &    -6.856 &     6.990 & \\
      5000 &       200 &     70.85 &   -22.461 &    16.869 & &     90.77 &    -4.000 &     8.271 & \\
      5000 &       300 &     67.53 &   -18.683 &    15.031 & &     92.25 &    -2.275 &     7.715 & \\
 \hline 
     10000 &        50 &     92.99 &    -7.200 &     7.317 & &     92.62 &    -1.846 &     3.309 & \\
     10000 &       100 &     96.31 &    -3.081 &     4.078 & &     98.15 &    -0.486 &     1.588 & \\
     10000 &       150 &     96.68 &    -2.018 &     3.242 & &     99.26 &    -0.192 &     0.937 & \\
     10000 &       200 &     97.79 &    -1.138 &     2.399 & &     98.89 &    -0.092 &     0.572 & \\
     10000 &       300 &     98.15 &    -0.463 &     1.585 & &     99.26 &    -0.001 &     0.456 & \\
 \hline
     20000 &        50 &     98.52 &    -0.001 &     2.468 & &     98.52 &     0.163 &     1.929 & \\
     20000 &       100 &     99.26 &     0.032 &     0.790 & &     99.63 &     0.041 &     0.799 & \\
     20000 &       150 &     99.26 &    -0.046 &     0.504 & &     99.26 &    -0.042 &     0.504 & \\
     20000 &       200 &    100.00 &     0.030 &     0.434 & &    100.00 &     0.030 &     0.435 & \\
     20000 &       300 &     99.63 &     0.036 &     0.283 & &     99.63 &     0.037 &     0.283 & \\
 \hline
     50000 &        50 &     99.26 &     0.013 &     1.400 & &     99.26 &     0.012 &     1.398 & \\
     50000 &       100 &     99.63 &     0.060 &     0.689 & &     99.63 &     0.060 &     0.691 & \\
     50000 &       150 &     99.63 &     0.053 &     0.441 & &     99.63 &     0.054 &     0.442 & \\
     50000 &       200 &     99.63 &     0.064 &     0.338 & &     99.63 &     0.064 &     0.335 & \\
     50000 &       300 &     99.63 &     0.069 &     0.269 & &     99.63 &     0.069 &     0.268 & \\
 \hline
    100000 &        50 &     99.26 &     0.182 &     1.279 & &     99.26 &     0.180 &     1.278 & \\
    100000 &       100 &    100.00 &     0.048 &     0.711 & &     99.63 &     0.055 &     0.771 & \\
    100000 &       150 &     99.26 &     0.094 &     0.533 & &     99.26 &     0.093 &     0.537 & \\
    100000 &       200 &     99.26 &     0.081 &     0.453 & &     99.26 &     0.082 &     0.451 & \\
    100000 &       300 &     99.26 &     0.097 &     0.362 & &     99.26 &     0.096 &     0.365 & \\

\enddata
\end{deluxetable}

\section{Summary}

We have developed a new software tool for automatic \ew\ measurement called \TAME\ for measuring \ews\ in a high-resolution spectrum. 
It has the following features: 
\begin{itemize}
 \item \TAME\ can automatically measure \ews\ for a large set of lines in a spectrum simultaneously.
 \item \TAME\ offers an interactive mode, in which a user can adjust the local continuum level precisely and change parameters such as the SMOOTHER, radial velocity, and type of fitting profile (Gaussian/Voigt).
 \item \TAME\ provides a text file including the \ews\ with a format suited for MOOG code and a graphical post-script file for confirming the \ew\ results when performing abundance analysis.
\end{itemize}

We verified \TAME\ in two ways. 
By using the solar spectrum, we measured solar \ews\ by \TAME\ and compared them with those obtained by the traditional method with IRAF {\tt splot} task. 
The \ews\ measured by \TAME\ showed a good agreement with the precise manual measurements made using IRAF, with a standard deviation of only 1.76 m\AA, and the atmospheric parameters of the Sun were determined to \teff\ $=$ 5791 K, \logg\ $=$ 4.54 dex, [Fe/H] = 0.03 dex, and $\xi_t = $ 0.81 \kms\ from the \ew\ result of \TAME. 
 
In order to examine the effect of the S/N ratio and spectral resolution on \ew\ measurement, we  performed \ew\ measurement for different synthetic spectra by using \TAME\ in the fully automatic mode without any manual interactions.
From the test results obtained for the synthetic spectra, we concluded that the \ew\ measurements obtained by \TAME\ are reliable for high-resolution spectra with R $\gtrsim$ 20000 and found that the errors in \ews\ could be expected to be less than 1 m\AA\ for a S/N ratio $\gtrsim$ 100.



\acknowledgments
This work was partially supported by Human Resources Development Program of the National Research Foundation of Korea and the WCU grant R31-10016 funded by the Korean Ministry of Education, Science, and Technology.


\end{document}